\documentclass{emulateapj}

\usepackage[normalem]{ulem}
\usepackage{color}

\newcommand{\sfedense}{\mbox{$\rm SFE_{dense}$}}

\newcommand{\fdense}{\mbox{$\rm f_{dense}$}}

\slugcomment{Accepted for Publication in ApJL}

\shorttitle{Dense Gas and Star Formation in M51}
\shortauthors{Bigiel et al.}

\begin{document}

\title{The EMPIRE Survey: Systematic Variations in the Dense Gas Fraction and Star Formation Efficiency from Full-Disk Mapping of M51}

\author{Frank Bigiel\altaffilmark{1}, Adam K. Leroy\altaffilmark{2}, Maria J. Jim\'enez-Donaire\altaffilmark{1}, J\'er\^ome Pety\altaffilmark{3,4}, Antonio Usero\altaffilmark{5}, Diane Cormier\altaffilmark{1}, Alberto Bolatto\altaffilmark{6}, Santiago Garcia-Burillo\altaffilmark{5}, Dario Colombo\altaffilmark{7}, Manuel Gonz\'alez-Garc\'ia\altaffilmark{8}, Annie Hughes\altaffilmark{9,10}, Amanda A. Kepley\altaffilmark{11}, Carsten Kramer\altaffilmark{12}, Karin Sandstrom\altaffilmark{13}, Eva Schinnerer\altaffilmark{14}, Andreas Schruba\altaffilmark{15}, Karl Schuster\altaffilmark{3}, Neven Tomicic\altaffilmark{14}, Laura Zschaechner\altaffilmark{14}}
\altaffiltext{1}{Institut f\"ur theoretische Astrophysik, Zentrum f\"ur Astronomie der Universit\"at Heidelberg, Albert-Ueberle Str. 2, 69120 Heidelberg, Germany; bigiel@uni-heidelberg.de}
\altaffiltext{2}{Department of Astronomy, The Ohio State University, 140 W 18$^{\rm th}$ St, Columbus, OH 43210, USA}
\altaffiltext{3}{Institut de Radioastronomie Millim\'etrique (IRAM), 300 Rue de la Piscine, F-38406 Saint Martin d'H\`eres, France}
\altaffiltext{4}{Observatoire de Paris, 61 Avenue de l'Observatoire, F-75014 Paris, France}
\altaffiltext{5}{Observatorio Astron\'omico Nacional, Alfonso XII 3, 28014, Madrid, Spain}
\altaffiltext{6}{Department of Astronomy and Laboratory for Millimeter-Wave Astronomy, University of Maryland, College Park, MD 20742, USA}
\altaffiltext{7}{Department of Physics, University of Alberta, 4-181 CCIS, Edmonton, AB T6G 2E1, Canada}
\altaffiltext{8}{Instituto de Astrof\'isica de Andaluc\'ia IAA-CSIC, Glorieta de la Astronom\'ia s/n, E-18008, Granada, Spain}
\altaffiltext{9}{CNRS, IRAP, 9 Av. colonel Roche, BP 44346, F-31028 Toulouse cedex 4, France}
\altaffiltext{10}{Universit\'{e} de Toulouse, UPS-OMP, IRAP, F-31028 Toulouse cedex 4, France}
\altaffiltext{11}{National Radio Astronomy Observatory, 520 Edgemont Road, Charlottesville, VA 22903-2475, USA}
\altaffiltext{12}{Instituto de Radioastronom�a Milim�trica (IRAM), Av. Divina Pastora 7, Nucleo Central, E-18012 Granada, Spain}
\altaffiltext{13}{Center for Astrophysics and Space Sciences, Department of Physics, University of California, San Diego, 9500 Gilman Drive, La Jolla, CA 92093, USA}
\altaffiltext{14}{Max-Planck-Institut f\"ur Astronomie, K\"onigstuhl 17, 69117 Heidelberg, Germany}
\altaffiltext{15}{Max-Planck-Institut f\"ur extraterrestrische Physik, Giessenbachstrasse 1, 85748 Garching, Germany}

\begin{abstract}
We present the first results from the EMPIRE survey, an IRAM large program that is mapping tracers of high density molecular gas across the disks of nine nearby star-forming galaxies. Here, we present new maps of the 3-mm transitions of HCN, HCO$^+$, and HNC across the whole disk of our pilot target, M51. As expected, dense gas correlates with tracers of recent star formation, filling the ``luminosity gap'' between Galactic cores and whole galaxies. In detail, we show that both the fraction of gas that is dense, \fdense\ traced by HCN/CO, and the rate at which dense gas forms stars, \sfedense\ traced by IR/HCN , depend on environment in the galaxy. The sense of the dependence is that high surface density, high molecular gas fraction regions of the galaxy show high dense gas fractions and low dense gas star formation efficiencies. This agrees with recent results for individual pointings by \citet{usero15} but using unbiased whole-galaxy maps. It also agrees qualitatively with the behavior observed contrasting our own Solar Neighborhood with the central regions of the Milky Way. The sense of the trends can be explained if the dense gas fraction tracks interstellar pressure but star formation occurs only in regions of high density contrast.
\end{abstract}


\section{Introduction}
\label{intro}

\begin{figure*}[t]
\epsscale{.7}
\plotone{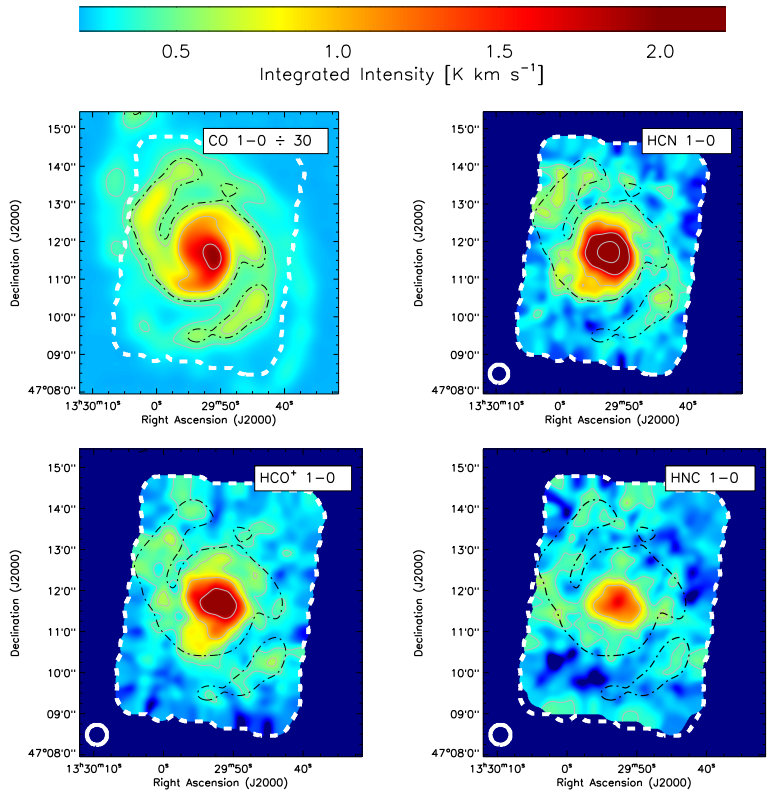}
\caption{Integrated intensity maps from our new survey of HCN$(1-0)$, HCO$^{+}$$(1-0)$ and HNC$(1-0)$ across M51. For reference, the upper left panel shows the PAWS $^{12}$CO$(1-0)$ single dish map, scaled down by a factor of $30$ to match the intensity scale of the other lines. Gray lines indicate contour levels of $I = 0.2, 0.4, 0.8, 1.6, 3.2$ and $6.4$\,K\,km\,s$^{-1}$. A black line shows the same CO contour ($I = 10$\,K\,km\,s$^{-1}$ in the unscaled map) in each panel. The beam size of 30$\arcsec$ is shown in the lower left corner. The dense gas tracers are particularly bright in the center and along the spiral arms. Their emission correlates well with the bulk molecular gas traced by CO emission.}
\label{fig1}
\end{figure*}

\begin{figure*}[t]
\epsscale{1.1}
\plottwo{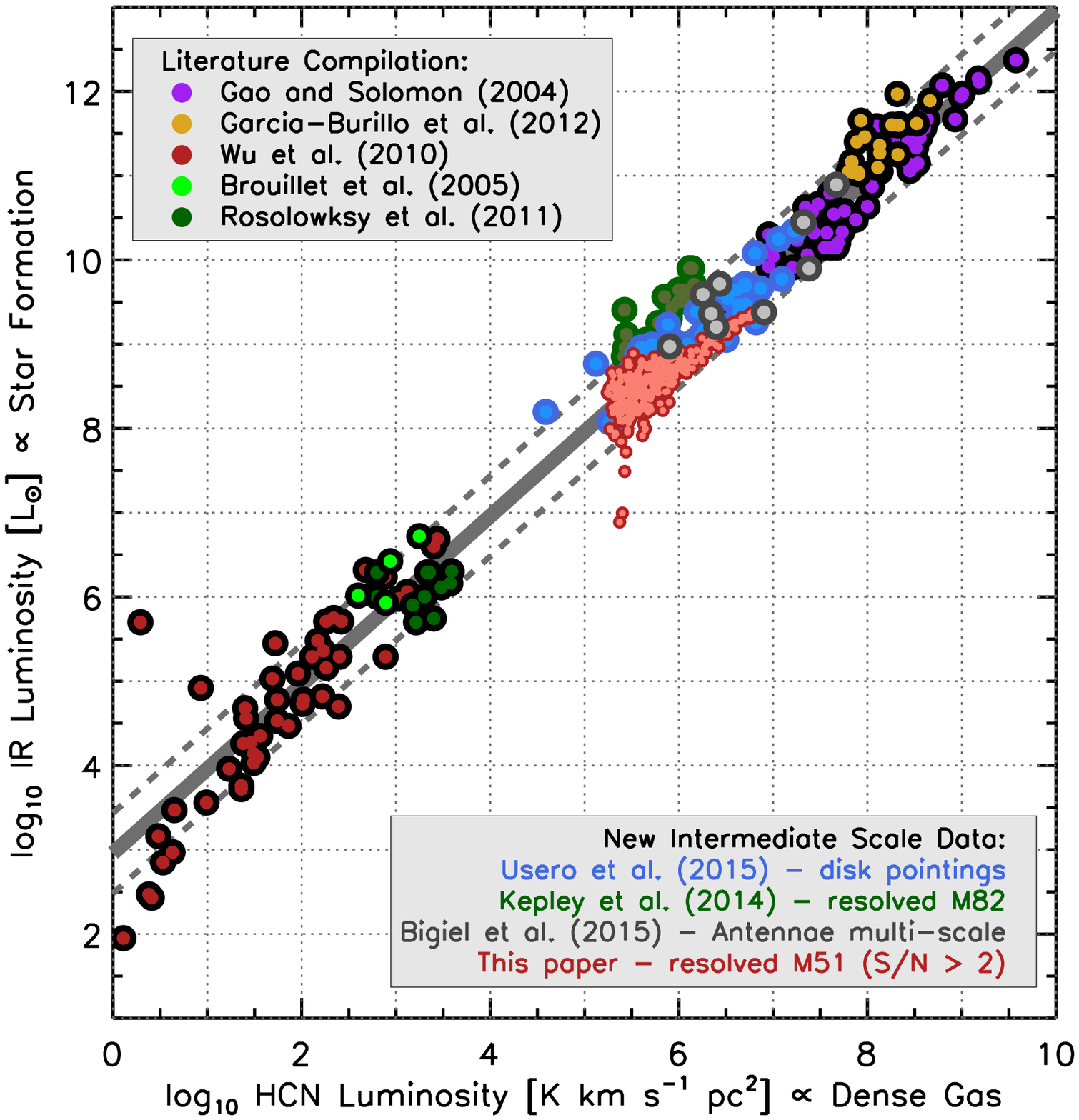}{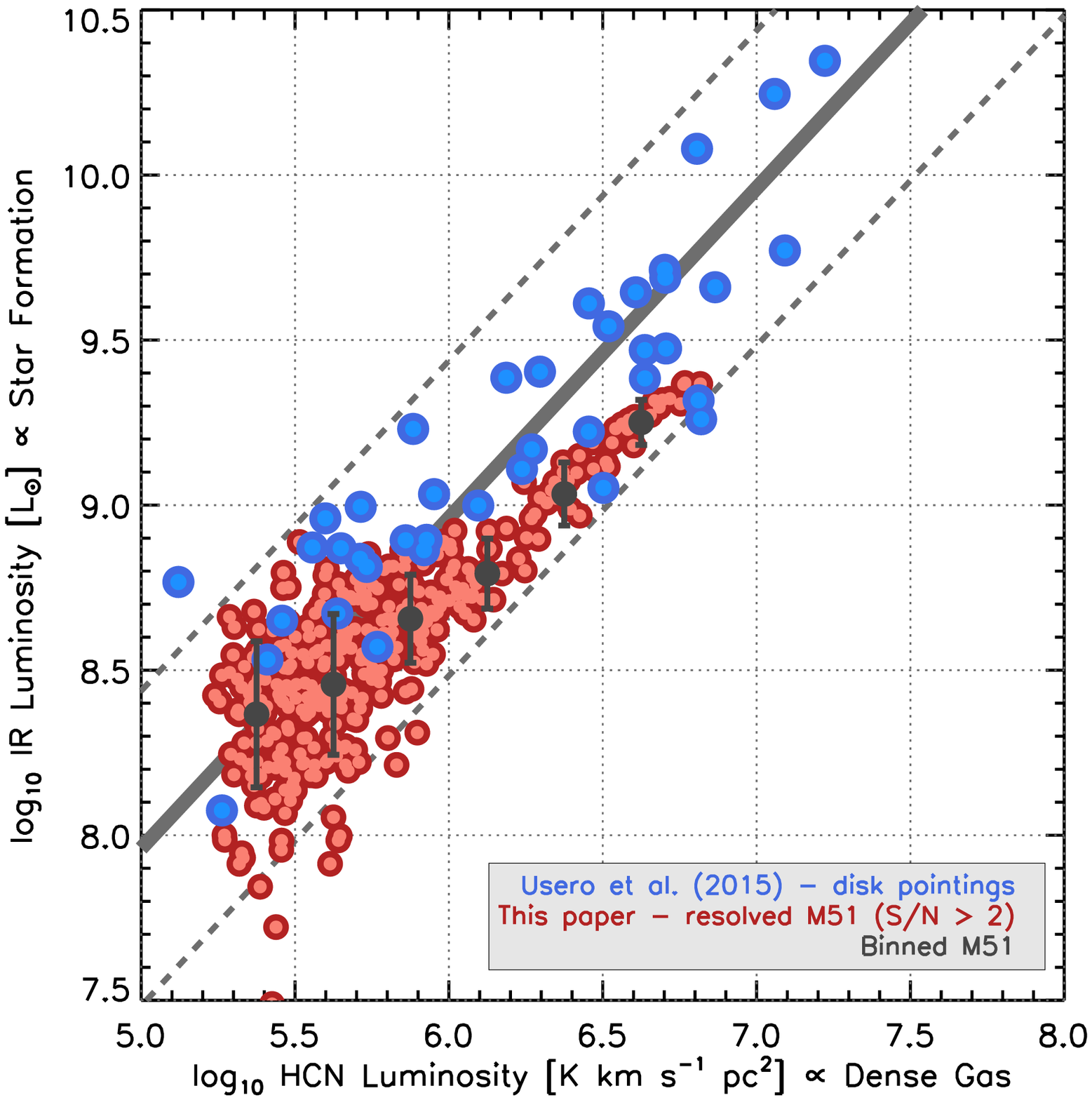}
\caption{{\em (left)} IR luminosity, tracing recent star formation, as a function of HCN luminosity, tracing dense gas, for structures from Milky Way cores and Local Group clouds to whole starburst galaxies; compilation from \citet{leroy15}. Along with other recent results, our new M51 data extend the observed correlation into the ``luminosity gap'' corresponding to large parts of galaxies. All data combined scatter about the ratio $L_{\rm TIR} / L_{\rm HCN} \approx 900$~L$_\odot$~(K~km~s$^{-1}$~pc$^2$)$^{-1}$ (gray line; dashed lines show a factor of 3 scatter). ({\em right}) Zoom-in on data for the full-galaxy map of M51 and the pointings from \citet{usero15}. The typical uncertainty on the HCN luminosity for each line of sight is $\sim10^5$K~km~s$^{-1}$~pc$^2$. IR correlates well with HCN luminosity for individual regions, but not all regions show the same IR-to-HCN ratio; both data sets show evidence for a mildly sub-linear slope \citep[see also][]{chen15}, indicating an environment dependent efficiency of star formation in dense gas.}
\label{fig2}
\end{figure*}

\begin{figure*}[t]
\epsscale{1.1}
\plottwo{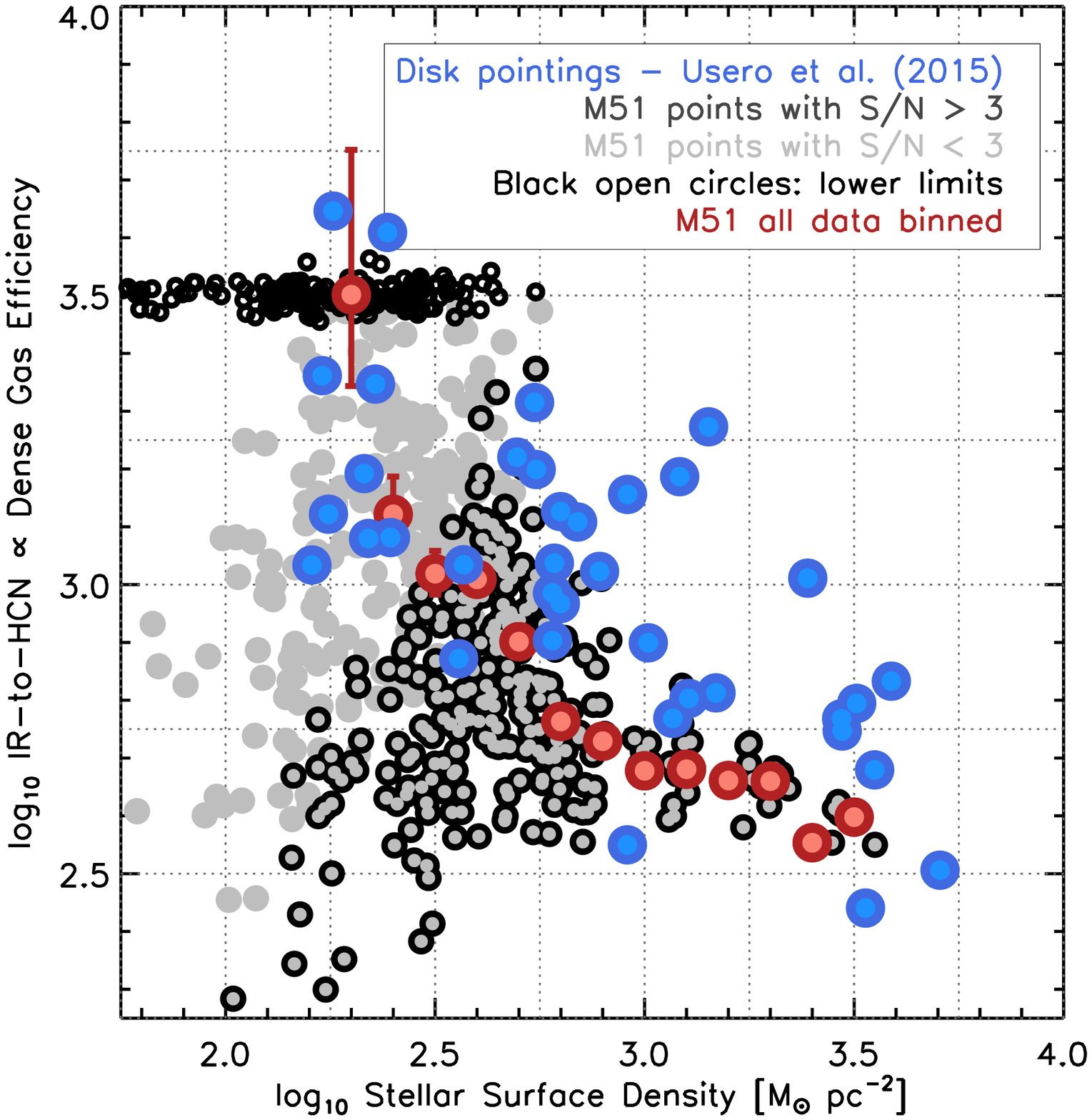}{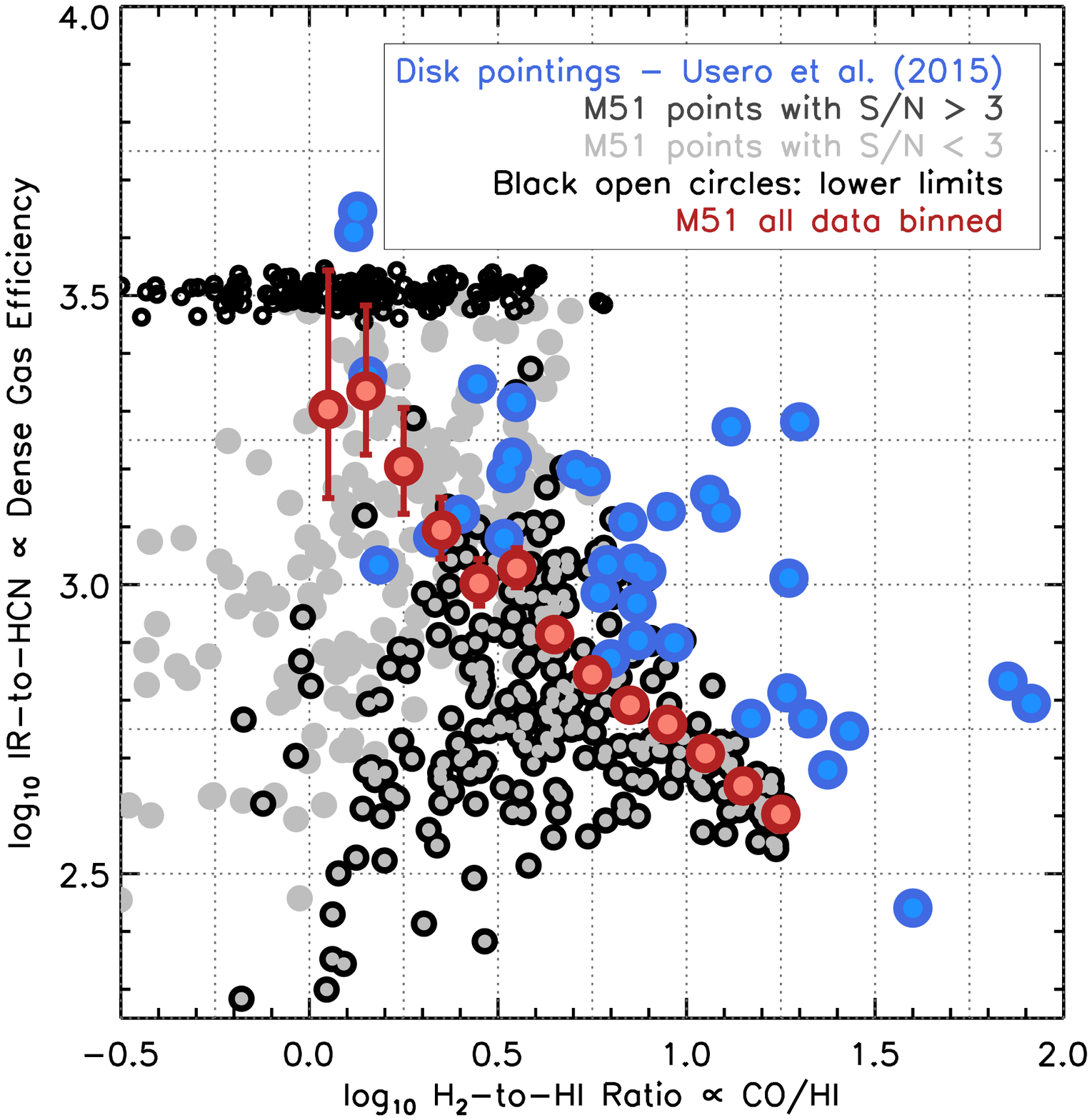}
\caption{IR-to-HCN ratio, tracing \sfedense\ as a function of ({\em left}) local stellar surface density, $\Sigma_\star$, and ({\em right}) molecular-to-atomic gas ratio, H$_2$/{\sc Hi}$\,\propto$ CO/{\sc Hi}. Gray points with black outlines show points in M51 where HCN is detected at S/N$>3$. Light gray points with no outline show points with lower S/N. Black open points show lower limits - that is, the measured IR/HCN value is above this value or negative (due to $I_{\rm HCN}<0$). A small amount of noise has been added to the $y$-value of the upper limits to distinguish points. Red points show the binned trend for the M51 data; including upper limits. Blue points show results for the \citet{usero15} sample. The disk pointings overlap results for our whole-galaxy map, though M51 has a somewhat lower IR-to-HCN ratio than other galaxies in the survey. Both data sets show that \sfedense\ appears to be lower in the high surface density, high molecular gas fraction, central, high pressure parts of galaxies.}
\vspace{0.2in}
\label{fig3}
\end{figure*}

\begin{figure*}[t]
\epsscale{1.1}
\plottwo{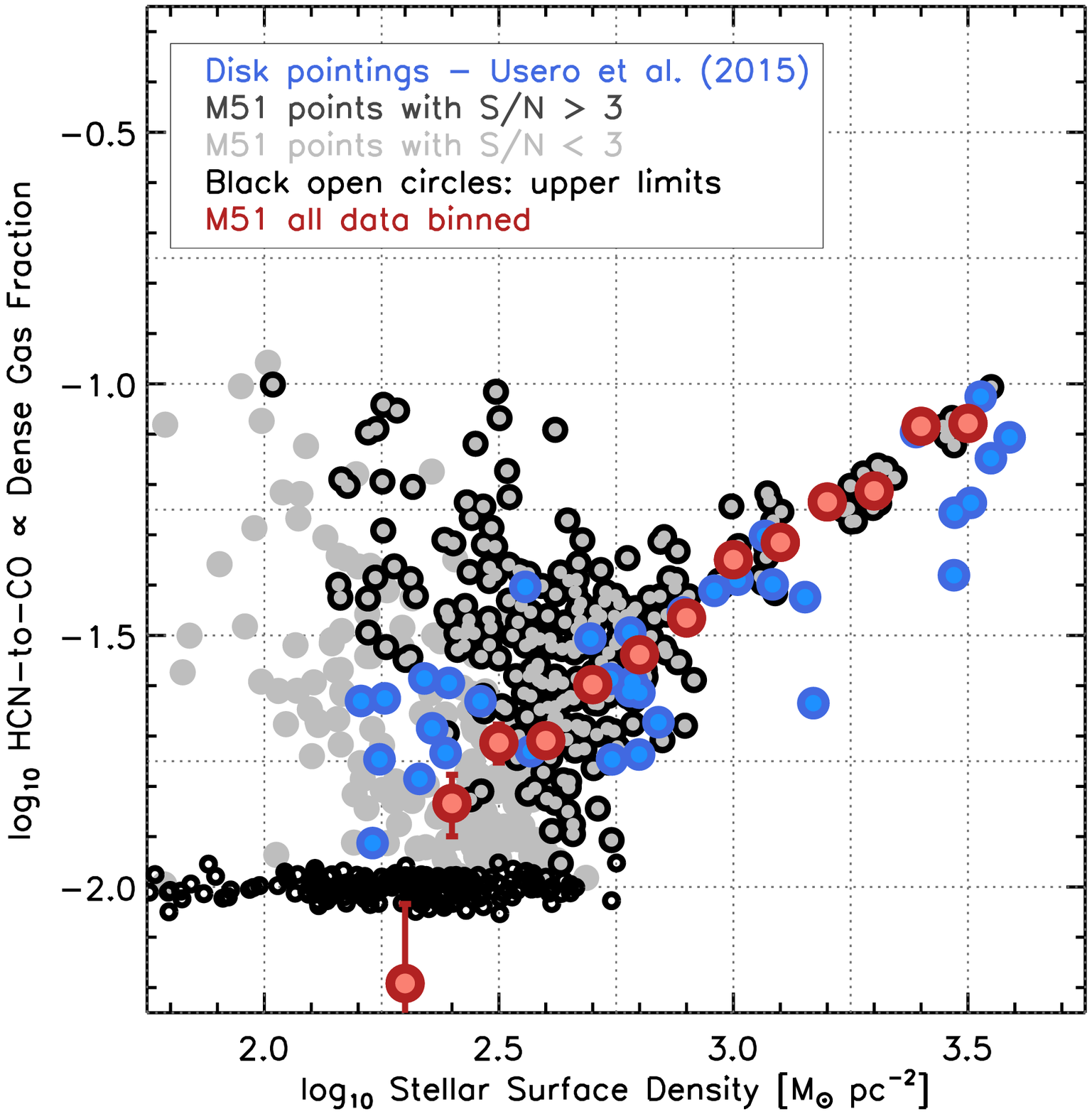}{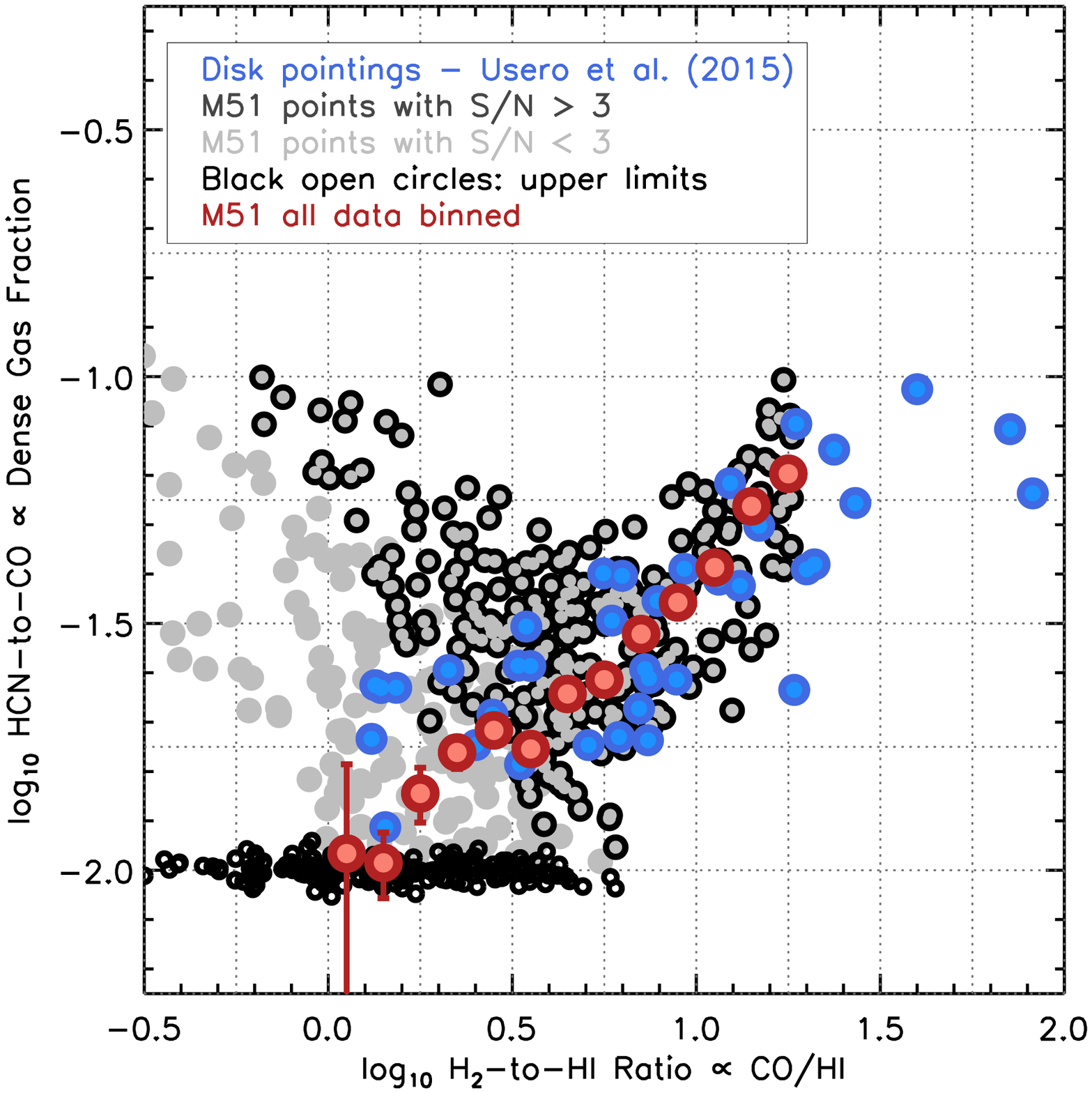}
\caption{HCN-to-CO ratio, tracing \fdense\ as a function of ({\em left}) local stellar surface density, $\Sigma_\star$, and ({\em right}) molecular-to-atomic gas ratio, H$_2$/{\sc Hi}\,$\propto$ CO/{\sc Hi}. For a description of the individual data points see caption of Figure \ref{fig3}. The disk pointings agree with our results for M51. Both data sets show \fdense\ to depend strongly on local conditions in a galaxy, with the sense of high dense gas fractions in the high surface density, heavily molecular, inner, high pressure regions of M51.}
\vspace{0.2in}
\label{fig4}
\end{figure*}

\begin{figure*}[t]
\epsscale{1.1}
\plotone{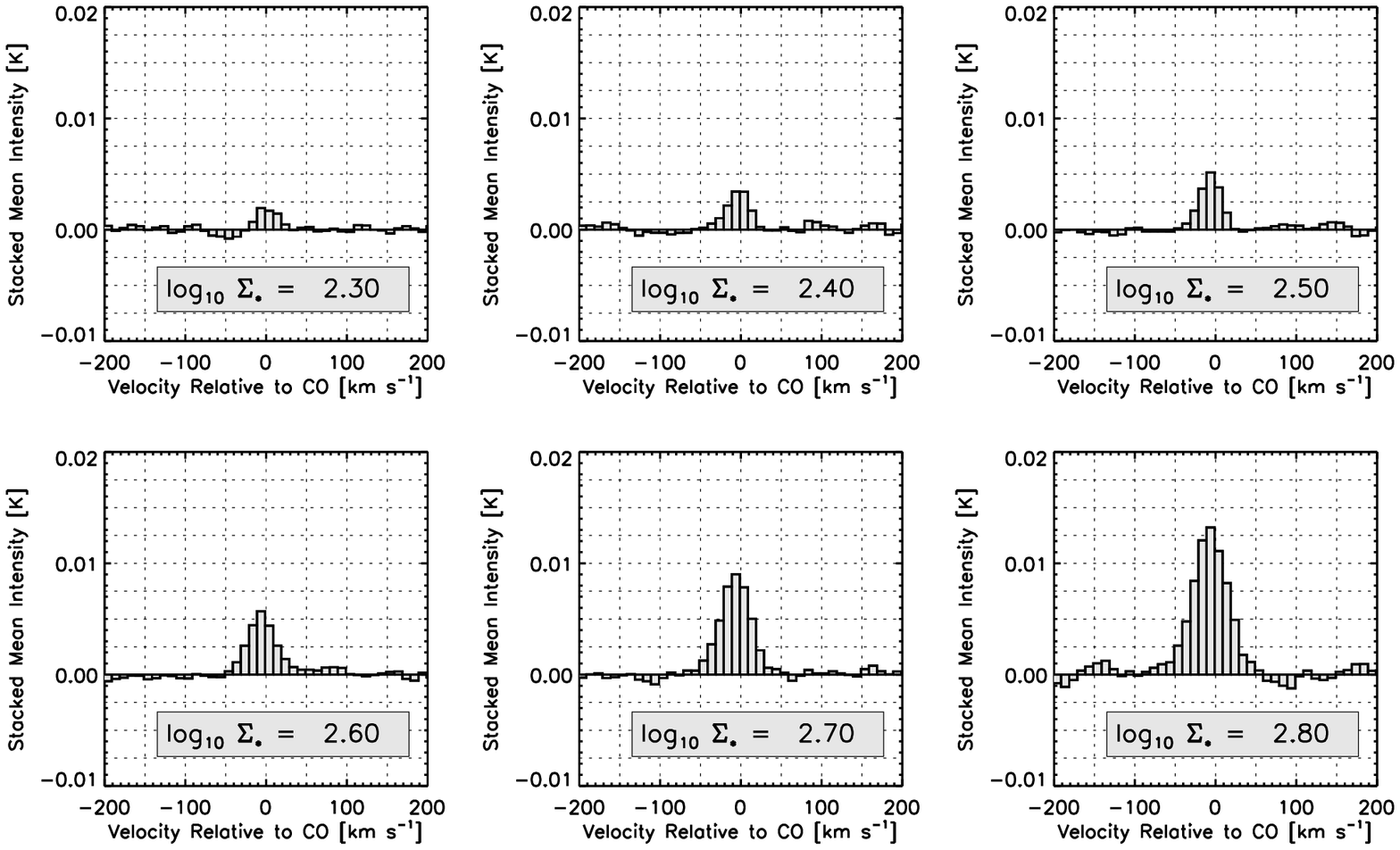}
\caption{Stacked HCN spectra for our lowest six stellar surface density bins in Figures \ref{fig3} and \ref{fig4}. In each panel, we sum all spectra in an 0.1~dex-wide bin centered at the indicated stellar surface density. To do this, we normalize all spectra by the local mean CO velocity and then average all spectra, including upper limits. Because we average over a large area in the low $\Sigma_\star$ bins, each bin yields a high significance detection of HCN intensity, despite the modest signal to noise in each individual beam.}
\vspace{0.2in}
\label{fig5}
\end{figure*}

Stars form out of dense molecular gas. This is evident from Milky Way studies that resolve individual clouds and isolate star-forming clumps \citep[e.g., see][]{lada03,heiderman10,lada10,andre14}. It can also be seen from the good correspondence between spectroscopic tracers of dense gas, like the low-$J$ transitions of HCN and HCO$^+$ \citep{gao04a,gao04b,gracia06,juneau09,garcia12,kepley14,usero15,chen15} or higher-J molecular lines \citep[e.g.,][]{zhang14,liu15} to tracers of recent star formation in star-forming galaxies and Milky Way cores \citep{wu05,wu10}. Thus, understanding what sets the equilibrium fraction of gas that is dense and how dense gas relates to star formation across the galaxy population is a crucial next step to understand how galaxies convert their gas reservoirs into stars.

Dense structures within clouds are very compact and thus hard to resolve at extragalactic distances. Spectroscopy of molecules with different density sensitivities offers the best way to systematically pursue this topic in other galaxies. By contrasting emission from lines excited at low density (e.g., low-J CO emission, ${\rm n_{eff}\approx10^2~cm^{-3}}$) with emission from lines excited only at high density (e.g. ${\rm HCN(1-0)~emission, n_{eff}\approx10^5~cm^{-3}}$), one can gauge the fraction of gas in a beam that is dense, $\fdense = {\rm M_{dense} / M_{gas}}$. Similarly, by comparing tracers of the recent star formation rate (SFR) to lines that trace the dense gas mass, one can explore variations in the efficiency with which dense gas forms stars, $\sfedense = {\rm SFR / M_{dense}}$.

With these goals in mind, we are using the IRAM 30-m telescope\footnote{Based on observations carried out with the IRAM 30m Telescope. IRAM is supported by INSU/CNRS (France), MPG (Germany) and IGN (Spain).} to carry out the first wide-area multi-line mapping survey targeting tracers of dense gas across the disks of star-forming galaxies. The ``EMIR Multiline Probe of the ISM Regulating Galaxy Evolution'' (EMPIRE, Bigiel et al., in prep.) is mapping a suite of density-sensitive transitions in the 3-mm atmospheric window over the full area of active star formation in nine nearby disk galaxies. This paper reports the first results of this survey for the prototype target, M51a (NGC 5194).

EMPIRE's main goals include understanding how \fdense\ and \sfedense\ depend on environment within a galaxy. The $\sim 30\arcsec$ beam of the IRAM 30-m, $\approx 1$--$2$~kpc at the distance of our targets, does not resolve individual clouds. However, it does localize quantities related to galactic structure like the stellar surface density, gas surface density, and --- to some degree -- dynamical environment, allowing one to measure how \fdense\ and \sfedense\ depend on these factors. 

EMPIRE builds on the seminal work by \citet{gao04a}, who surveyed HCN from the bright regions of active galaxies to show that $\sfedense$ varies weakly among galaxies, while $\fdense$ varies strongly. This provided a main piece of observational evidence for a universal gas density threshold for star formation. Surveying similar systems and using the HCO$^+$ line, \citet{garcia12} showed that this situation appears more complex, with $\sfedense$ varying between starburst and normal star-forming galaxies. These surveys targeted whole bright galaxies. For individual pointings across 30 disk galaxies spanning a range of physical conditions, \citet{usero15} showed that both \fdense\ and \sfedense\ within their $\sim 1$--$2$~kpc beam depended strongly on these conditions. 
 
Given the apparent dependence of \fdense\ and \sfedense\ on local conditions, the natural next step must be complete, unbiased maps of a set of galaxies with a wide range of global and local conditions. As the first systematic multi-line mapping survey targeting the whole disks of galaxies, EMPIRE represents this natural next step. This letter demonstrates from complete mapping that both \fdense\ and \sfedense\ vary systematically across the disk of M51.

\section{Observations}
\label{data}

We observed M51 at the IRAM 30-m telescope for 75 hours in seven consecutive runs in July and August 2012 under average summer conditions. We used the 3\,mm band of the sideband-separating dual-polarization EMIR receiver \citep{carter12} and the Fourier transform spectrometers. This yields a bandwidth of 15.6\,GHz per polarization with a channel spacing of 195\,kHz. We tuned to simultaneously measure the ${\rm HCO^+(1-0) [89.19\,GHz]}$, ${\rm HCN(1-0) [88.63 GHz]}$, ${\rm HNC(1-0) [90.66 GHz]}$, H$^{13}$CO$^+(1-0)$ [86.75 GHz], H$^{13}$CN$(1-0)$ [86.34 GHz], and HN$^{13}$C$(1-0)$ [87.09 GHz] transitions. This letter reports first results for HCN emission. Jim\'enez-Donaire et al. (in prep.) present the optically thin isotopologues and report their ratios to other lines.

M51 was mapped in on-the-fly mode, with a dump time of 0.5\,s and a scanning speed of $9\arcsec/$s, which yielded 6 dumps per beam along the scanning direction at the $\approx 28\arcsec$ angular resolution of the ${\rm HCO^+(1-0)}$ line. We covered a field-of-view of $\sim4.2\arcmin\times5.7\arcmin$ at a position angle of $-7.5\degr$ centered on ($\alpha={\rm 13^h29^m52.532^s}$, $\delta={\rm 47^d11^m41.98^s}$). The top left panel of Figure \ref{fig1} shows the footprint of the map on top of the PAWS $^{12}$CO$(1-0)$ map \citep{schinnerer13,pety13}, demonstrating that we cover most of the area of bright CO emission in the galaxy.

We reduced the data using the \textsc{gildas/class} software\footnote{\texttt{http://www.iram.fr/IRAMFR/GILDAS}; for more information see \citet{pety05}.}. First, we extracted a 300\,MHz-wide window around the rest frequency of each line. We flagged spectra with noise larger than expected from the radiometer equation. The rest of the spectra were gridded into a data cube using a Gaussian kernel with a FWHM one third of the telescope beam and a channel width of 7~km~s$^{-1}$. We fit and removed a third order baseline using the signal-free part of each spectrum (determined by comparison to $^{12}$CO emission) in the cube.

We convolve all data to $30\arcsec \approx 1.1$~kpc resolution at the $7.6$~Mpc distance to M51 \citep{ciardullo02} and adopt an inclination of $i\approx22\deg$. We sample all maps using a common set of hexagonally-packed points spaced by $15\arcsec$, or one half beam. For each point, we calculate intensities integrated over the range of velocities with bright $^{12}$CO$(1-0)$ emission. Uncertainties in the integrated intensities are estimated using the noise estimated from the signal-free region and the width ($\sim40$~km~s$^{-1}$) of the integration window. A typical uncertainty for an individual $30\arcsec$ line of sight in the EMPIRE maps is 0.06~K~km~s$^{-1}$. 

Note that we do not sigma clip the data or otherwise restrict our measurements to regions of bright HCN emission. We calculate an HCN intensity with an associated uncertainty for every line of sight in the map. In our presentation of results below, we also average these individual lines of sight over large areas (by binning HCN intensity sorted by, e.g., stellar surface density), which yields a signal at high significance. This has the large advantage that our results remain sensitive to faint emission that may not be detected at high significance in individual lines of sight. We rigorously propagate and plot uncertainties in the mean trends, accounting for the oversampling of our data.

In addition to the EMPIRE data, we use the IRAM 30-m map of $^{12}$CO$(1-0)$ emission from PAWS and multi-band {\em Herschel} imaging \citep{mentuch12}, which we combine to estimate the total infrared surface brightness along the line of sight following \citet{galametz13}. We also use the THINGS {\sc Hi} map from \citet{walter08} and a map of stellar surface density estimated from {\em Spitzer} near-IR photometry by \citet{meidt12,querejeta15} using a 3.6$\mu$m mass-to-light ratio of ${\rm \sim 0.5 M_\odot / L_\odot}$, so that ${\rm\Sigma_\star \left[ M_\odot~{pc}^{-2}\right] = 350\,I_{3.6,corr} \,\left[ {MJy~sr}^{-1}\right]}$. The {\em Herschel} and {\em Spitzer} data are extremely high signal to noise, so that the dominant uncertainty is the calibration. This is uncertain at the $\sim 5\%$ level, but systematic. Similarly, the CO and {\sc Hi} data have much higher signal-to-noise than EMPIRE data. The dominant statistical uncertainty throughout the paper is thus that of the HCN intensity; we propagate this and show it in the figures and mean trends.

\section{Results}
\label{results}

Figure \ref{fig1} shows the integrated intensity maps of HCN~(1-0), HCO$^+$~(1-0) and HNC~(1-0), the three main dense gas tracers in EMPIRE. For comparison, the top left panel also shows the $^{12}$CO~(1-0) map from PAWS \citep{pety13}, tracing the overall distribution of molecular gas. We detect emission from all three dense gas tracers across the disk of M51, with emission brightest in the center and along the spiral arms. All three dense gas tracers align well with the structure of the CO emission, with HCN moderately brighter than HCO$^+$, and both brighter than HNC. CO is far brighter than all three molecules; the map that we show has been scaled by the typical HCN-to-CO ratio of 1-to-30 to bring it onto a common intensity scale with the EMPIRE dense gas maps.

Figure \ref{fig2} compares the amount of dense gas, traced by HCN, to the amount of recent star formation, traced by IR luminosity. In the left panel, we convert each individual sampling point to a luminosity and plot these on the global scaling relation between IR luminosity and HCN luminosity following \citet{gao04b} and \citet{wu05}. The figure shows that our M51 data extend the IR-HCN correlation seen for whole galaxies to much lower luminosities, overlapping work for individual pointings by \citet{usero15} and \citet{bigiel15} and maps by \citet[][M82; see also Chen et al. (2015), M51]{kepley14}. These data help to fill in the ``luminosity gap'' between individual Galactic cores \citep{wu10} or clouds \citep{rosolowsky11,brouillet05} and whole active star-forming galaxies \citep{gao04b,garcia12}. Across $\sim 8$ orders of magnitude in luminosity, the data scatter around roughly the same IR-to-HCN ratio, $L_{\rm TIR} / L_{\rm HCN} \approx 900$~L$_\odot$~(K~km~s$^{-1}$)$^{-1}$ \citep[][see also Gao et al. 2007]{gao04b}, shown as a thick gray line.

The right panel zooms in on the luminosity-luminosity relationship for only our new M51 data and the disk pointings of \citet{usero15}. Here each point shows an individual $\sim$ kpc sized beam (sampling point) with the black points and error bars showing a running median and the $1\sigma$ scatter in bins of fixed HCN luminosity. In agreement with the recent work of \citet{chen15}, there is a good correspondence between star formation (IR) and dense gas (HCN) across the disk of M51, with a Spearman rank correlation coefficient of 0.8 relating the two. At a given HCN luminosity, the $1\sigma$ scatter is between a factor of 1.3 at high luminosities and 2 near the low luminosity end. M51 shows similar behavior to the pointings from \citet{usero15}, though it is somewhat brighter in HCN than the other galaxies in the sample at intermediate luminosities. This might be a result of the ongoing merger driving gas in M51 to higher densities but also adding turbulence that prevents it from collapsing. In any case, we expect that comparison with the other full-galaxy maps from EMPIRE will illuminate the origin of this offset.

The right panel in Figure \ref{fig2} also shows that across the disk of M51, the relationship of IR to HCN is not perfectly 1-to-1, showing a high degree of scatter and a mildly sublinear slope. That is, \sfedense\ as traced by the IR-to-HCN ratio (fixed for diagonal lines) is lower at high luminosities in both our M51 map (by about 60$\%$) and the \citet{usero15} sample. \citet{usero15} found \sfedense\ to depend on environment, being lower at high stellar surface densities and high molecular fractions \citet{chen15} observed qualitatively similar behavior in their analysis of M51, noting that the IR-to-HCN ratio drops with decreasing radius in the galaxy.

In Figure \ref{fig3} we show that the same is true for our whole-galaxy map of M51. The IR-to-HCN ratio varies systematically across the galaxy, becoming lower in regions with high stellar surface densities and high molecular fractions. There is good quantitative agreement with the individual \citet{usero15} pointings for $30$ galaxies. The trend is particularly evident in the binned relation (red); here the error bars reflect propagated statistical uncertainty in the mean, which is dominated by statistical noise in the HCN data.

The amount of dense gas also varies systematically across M51, showing trends opposite to the sense of what we see for \sfedense. Figure \ref{fig4} shows that the fraction of gas mass in a dense phase increases with increasing stellar surface density and molecular gas fraction, so that at lower galactocentric radii and higher interstellar pressures, more of the gas is dense. Again, the binned (red) trend for M51 shows continuous variation at high significance. This agrees with the scenario proposed by \citet{helfer97} based on early observations of HCN in the inner regions of a few galaxies, it is quantitatively consistent with the observed results from \citet{usero15} and qualitatively consistent with what is seen in our own Galaxy \citep[e.g.,][]{longmore13}. It presents a clear challenge to the idea that star formation proceeds in a universal way above a gas density threshold similar to the effective density of ${\rm HCN(1-0)}$.

The very significant trends in the red (binned) points in Figures \ref{fig3} and \ref{fig4} emerge from a large set of individual, lower signal-to-noise pointings. Using spectral stacking techniques similar to \citet{schruba11}, we have verified that these averaged HCN measurements reflect an astronomical signal; that is, after averaging it comes from a spectral line feature coincident in velocity with CO. Figure \ref{fig5} shows the results of this stacking for HCN emission from the six lowest stellar surface density bins in Figures \ref{fig3} and \ref{fig4}. In each case, the stacking yields a clear, significant astrophysical line signature. We emphasize that this averaging approach is fundamentally similar to what is done to obtain a deep integration with any single dish telescope. In fact, such deep integrations on individual sparse pointings were carried out by \citet{usero15}. We show here that the two approaches yield consistent results.

Both stellar surface density and molecular gas fraction correlate with interstellar pressure and anti-correlate with galactocentric radius, so that in M51 \sfedense\ appears to be lowest and \fdense\ appears to be highest in the high pressure, high surface density regions near the center of the galaxy. Qualitatively, the environment dependence of \fdense\ that we observe may be explained if most of the molecular gas is in approximate equilibrium with the hydrostatic pressure in the galaxy \citep[e.g.,][]{helfer97,hughes13}. In this case, the higher stellar surface densities and molecular fractions indicate higher interstellar gas pressures, which in turn shifts the overall gas density distribution to higher values \citep[e.g.,][]{elmegreen93,blitz06,ostriker10,shi11}. This leads a larger fraction of the gas to be dense. The changing \sfedense\ may then be explained if star formation only occurs in regions that show a high density contrast relative to the local mean gas density, a scenario consistent with current turbulent models of star formation \citep[e.g., see][for more discussion]{federrath12,usero15}. Then in the dense parts of galaxies, the average gas cloud may be denser and better at emitting HCN, but the HCN-emitting gas may also no longer correspond to the high density, self-gravitating, immediately star-forming tail of the density distribution. 

In forthcoming papers, we test this and competing scenarios using the entire sample of nine EMPIRE galaxies and our full suite of density-sensitive molecular lines. Doing so, we expect to make the most thorough test to date of how physical conditions in the molecular gas depend on galactic environment and, in turn, influence star formation.

\section{Summary}

We present new, sensitive maps of M51 in lines sensitive to high density gas. These represent the first result of the IRAM large program ``EMPIRE,'' which aims to use multi-line spectroscopy to relate physical conditions in the cold ISM to star formation in $9$ nearby galaxies. Compared to previous work targeting whole galaxies, we confirm a tight scaling between the amount of dense gas, traced by HCN, and the recent star formation rate, traced by IR emission. But we show that the underlying physics are more complex: we show that both the dense gas fraction and the star formation efficiency of dense gas appear to depend on local conditions in the disk of M51. The sense of this trend is a high \fdense\ and a low \sfedense\ in the high surface density, high molecular fraction parts of the disk. This agrees with work on individual scattered pointings in $30$ galaxy disks by \citet{usero15} but here the results have no bias, they cover the entire galaxy, and by binning the data we show these relations at very high statistical significance. The result is consistent with the ISM pressure playing an important role in setting the density of the ISM and the ability of dense gas to form stars.\\

We thank Ga\"elle Dumas for assisting with data reduction and Fabian Walter for helpful discussions and feedback on the draft. FB, MJ and DC acknowledge support from DFG grant BI 1546/1-1. AH acknowledges support from the Centre National d'Etudes Spatiales (CNES). AU acknowledges support from Spanish MINECO grants AYA2012-32295 and FIS2012-32096. JP thanks the CNRS/INSU programme PCMI for support. NT acknowledges funding from DFG grant SCHI 536/8-2 as part of the priority program SPP 1573 ``Physics of the Interstellar Medium''.


\begin{thebibliography}{}

\bibitem[Andr{\'e} et al.(2014)]{andre14} Andr{\'e}, P., Di Francesco, J., Ward-Thompson, D., et al.\ 2014, Protostars and Planets VI, 27 
\bibitem[Bigiel et al.(2015)]{bigiel15} Bigiel, F., Leroy, A.~K., Blitz, L., et al.\ 2015, \apj, 815, 103 
\bibitem[Blitz \& Rosolowsky(2006)]{blitz06} Blitz, L., \& Rosolowsky, E.\ 2006, \apj, 650, 933 
\bibitem[Brouillet et al.(2005)]{brouillet05} Brouillet, N., Muller, S., Herpin, F., Braine, J., \& Jacq, T.\ 2005, \aap, 429, 153 
\bibitem[Carter et al.(2012)]{carter12} Carter, M., Lazareff, B., Maier, D., et al.\ 2012, \aap, 538, A89 
\bibitem[Chen et al.(2015)]{chen15} Chen, H., Gao, Y., Braine, J., \& Gu, Q.\ 2015, \apj, 810, 140 
\bibitem[Ciardullo et al.(2002)]{ciardullo02} Ciardullo, R., Feldmeier, J.~J., Jacoby, G.~H., et al.\ 2002, \apj, 577, 31 
\bibitem[Elmegreen(1993)]{elmegreen93} Elmegreen, B.~G.\ 1993, \apj, 411, 170  
\bibitem[Federrath \& Klessen(2012)]{federrath12} Federrath, C., \& Klessen, R.~S.\ 2012, \apj, 761, 156 
\bibitem[Galametz et al.(2013)]{galametz13} Galametz, M., Kennicutt, R.~C., Calzetti, D., et al.\ 2013, \mnras, 431, 1956 
\bibitem[Gao \& Solomon(2004a)]{gao04a} Gao, Y., \& Solomon, P.~M.\ 2004, \apjs, 152, 63 
\bibitem[Gao \& Solomon(2004b)]{gao04b} Gao, Y., \& Solomon, P.~M.\ 2004, \apj, 606, 271 
\bibitem[Gao et al.(2007)]{gao07} Gao, Y., Carilli, C.~L., Solomon, P.~M., \& Vanden Bout, P.~A.\ 2007, \apjl, 660, L93 
\bibitem[Graci{\'a}-Carpio et al.(2006)]{gracia06} Graci{\'a}-Carpio, J., Garc{\'{\i}}a-Burillo, S., Planesas, P., \& Colina, L.\ 2006, \apjl, 640, L135 
\bibitem[Garc{\'{\i}}a-Burillo et al.(2012)]{garcia12} Garc{\'{\i}}a-Burillo, S., Usero, A., Alonso-Herrero, A., et al.\ 2012, \aap, 539, A8
\bibitem[Heiderman et al.(2010)]{heiderman10} Heiderman, A., Evans, N.~J., II, Allen, L.~E., Huard, T., \& Heyer, M.\ 2010, \apj, 723, 1019 
\bibitem[Helfer \& Blitz(1997)]{helfer97} Helfer, T.~T., \& Blitz, L.\ 1997, \apj, 478, 162 
\bibitem[Hughes et al.(2013)]{hughes13} Hughes, A., Meidt, S.~E., Colombo, D., et al.\ 2013, \apj, 779, 46 
\bibitem[Juneau et al.(2009)]{juneau09} Juneau, S., Narayanan, D.~T., Moustakas, J., et al.\ 2009, \apj, 707, 1217
\bibitem[Kepley et al.(2014)]{kepley14} Kepley, A.~A., Leroy, A.~K., Frayer, D., et al.\ 2014, \apjl, 780, L13   
\bibitem[Lada \& Lada(2003)]{lada03} Lada, C.~J., \& Lada, E.~A.\ 2003, \araa, 41, 57 
\bibitem[Lada et al.(2010)]{lada10} Lada, C.~J., Lombardi, M., \& Alves, J.~F.\ 2010, \apj, 724, 687 
\bibitem[Lada et al.(2012)]{lada12} Lada, C.~J., Forbrich, J., Lombardi, M., \& Alves, J.~F.\ 2012, \apj, 745, 190 
\bibitem[Leroy et al.(2015)]{leroy15} Leroy, A.~K., Bolatto,  A.~D., Ostriker, E.~C., et al.\ 2015, \apj, 801, 25 
\bibitem[Liu et al.(2015)]{liu15} Liu, L., Gao, Y., \& Greve, T.~R.\ 2015, \apj, 805, 31 
\bibitem[Longmore et al.(2013)]{longmore13} Longmore, S.~N., Bally, J., Testi, L., et al.\ 2013, \mnras, 429, 987 
\bibitem[Mangum et al.(2007)]{mangum07} Mangum, J.~G., Emerson, D.~T., \& Greisen, E.~W.\ 2007, \aap, 474, 679 
\bibitem[Meidt et al.(2012)]{meidt12} Meidt, S.~E., Schinnerer, E., Knapen, J.~H., et al.\ 2012, \apj, 744, 17 
\bibitem[Mentuch Cooper et al.(2012)]{mentuch12} Mentuch Cooper, E., Wilson, C.~D., Foyle, K., et al.\ 2012, \apj, 755, 165
\bibitem[Ostriker et al.(2010)]{ostriker10} Ostriker, E.~C., McKee, C.~F., \& Leroy, A.~K.\ 2010, \apj, 721, 975  
\bibitem[Pety(2005)]{pety05} Pety, J.\ 2005, SF2A-2005:Semaine de l'Astrophysique Francaise, 721 
\bibitem[Pety et al.(2013)]{pety13} Pety, J., Schinnerer, E., Leroy, A.~K., et al.\ 2013, \apj, 779, 43 
\bibitem[Querejeta et al.(2015)]{querejeta15} Querejeta, M., Meidt, S.~E., Schinnerer, E., et al.\ 2015, \apjs, 219, 5 
\bibitem[Rosolowsky et al.(2011)]{rosolowsky11} Rosolowsky, E., Pineda, J.~E., \& Gao, Y.\ 2011, \mnras, 415, 1977 
\bibitem[Schinnerer et al.(2013)]{schinnerer13} Schinnerer, E., Meidt, S.~E., Pety, J., et al.\ 2013, \apj, 779, 42 
\bibitem[Schruba et al.(2011)]{schruba11} Schruba, A., Leroy, A.~K., Walter, F., et al.\ 2011, \aj, 142, 37
\bibitem[Shi et al.(2011)]{shi11} Shi, Y., Helou, G., Yan, L., et al.\ 2011, \apj, 733, 87 
\bibitem[Usero et al.(2015)]{usero15} Usero, A., Leroy, A.~K., Walter, F., et al.\ 2015, \aj, 150, 115 
\bibitem[Walter et al.(2008)]{walter08} Walter, F., Brinks, E., de Blok, W.~J.~G., et al.\ 2008, \aj, 136, 2563-2647 
\bibitem[Wu et al.(2005)]{wu05} Wu, J., Evans, N.~J., II, Gao, Y., et al.\ 2005, \apjl, 635, L173 
\bibitem[Wu et al.(2010)]{wu10} Wu, J., Evans, N.~J., II, Shirley, Y.~L., \& Knez, C.\ 2010, \apjs, 188, 313 
\bibitem[Zhang et al.(2014)]{zhang14} Zhang, Z.-Y., Gao, Y., Henkel, C., et al.\ 2014, \apjl, 784, L31 

\end{thebibliography}
\end{document}